# Silicon nanowire based exclusive-OR gate using nonlinear optics for 40Gb/s DPSK signals

F. Li,[1] T.D. Vo,[1] C. Husko,[1] M. Pelusi,[1] D-X. Xu,[2] A. Densmore,[2] R. Ma,[2] S. Janz,[2] B.J. Eggleton[1] and D.J. Moss[1]

[1]*Centre for Ultrahigh-bandwidth Devices for Optical Systems (CUDOS), Institute of Photonics and Optical Science (IPOS), School of Physics, The University of Sydney, NSW 2006, Australia*
[2]*Institute for Microstructural Sciences, National Research Council (NRC-CNRC), Ottawa, ON, Canada K1A-0R6*

**Abstract:** We demonstrate an all-optical XOR logic function for 40Gb/s differential phase-shift keyed (DPSK) data signals in the C-band, based on four-wave mixing (FWM) in a silicon nanowire. Error-free operation with a system penalty of ~ 3.0dB and ~ 4.3dB at $10^{-9}$ BER is achieved.

## 1. Introduction

All-optical nonlinear signal processing is seen [1,2] as a key for future telecommunication networks to overcome the electronic bandwidth bottlenecks as systems evolve towards 640Gb/s [3], Terabit Ethernet [4], and beyond [5,6]. All-optical logic gate functions, an important aspect to this, have been investigated in a variety of platforms for on-off keying (OOK) formats [7,8]. Recently, advanced modulation format signals involving phase-encoding, such as differential phase shift keying (DPSK), have attracted attention because of their high tolerance to system impairments and nonlinearities. All-optical logic gates for DPSK data have been successfully demonstrated in a number of platforms such as highly nonlinear silica fiber (HNLF) [9], semiconductor optical amplifiers (SOAs) [10], and periodically poled lithium niobate (PPLN) [11]. However, these platforms all suffer from drawbacks of one form or another, such as fiber dispersion in HNLF [9], free-carrier induced patterning effects in SOAs [10] that ultimately limit high-speed operation, and the requirement for precise temperature control in PPLN [12]. Therefore, alternative platforms are of significant interest to circumvent these issues and allow the integration of these devices into real systems.

Integrated nanophotonic devices, particularly in highly nonlinear materials, have been a key approach to reducing operating power requirements of all-optical devices by increasing the nonlinear parameter, $\gamma = \omega\, n_2 / c\, A_{eff}$ (where $A_{eff}$ is the waveguide effective area). Both silicon and chalcogenide nanowires have achieved extraordinarily high $\gamma$ 's ranging from 15 $W^{-1} m^{-1}$ in ChG waveguides [12] to 95 $W^{-1} m^{-1}$ in ChG tapered fiber nanowires [13], to 300 $W^{-1} m^{-1}$ in Si nanowires [14]. This led to the first demonstration of chip-based all-optical exclusive-OR (XOR) logic gate for 40Gb/s and 160Gb/s DPSK signals in ChG nanowires [15]. It would be of significant interest to be able to accomplish all-optical logic at these high bit rates in silicon nanophotonic devices, particularly since they are seen as being compatible with CMOS electronic circuits.

All-optical logic gates have been demonstrated in a number of silicon devices based on single [16] or cascaded [17] microring resonators, in waveguides using two-photon absorption [18,19], Raman gain [20], and interference effects [21,22] in couplers. However, the data rates in all of these demonstrations have been limited to < 1 Gb/s for on-off key signals, for a variety of reasons including the bandwidth limitation posed by ring resonators, even though silicon itself has been shown to be capable of all-optical signal processing at much higher speeds [23 - 26]. Third-order nonlinear processes such as parametric gain and four wave mixing [27] are particularly suited to all-optical signal processing where both maintaining the integrity of the signal phase while performing phase sensitive operations, is critical. This has been the primary motivation behind the recent work on all-optical phase amplification and regeneration [27, 28].

In this paper, we report the first high bit rate operation for all-optical signal processing in a silicon device. We demonstrate all-optical XOR logic operation on a 40Gb/s (33% Return-to-Zero) DPSK PRBS data stream in the C-band via non-degenerate four wave mixing (FWM) in a silicon nanowire. Our method is based on an approach proposed and demonstrated in HNFL [9] and chalcogenide glass waveguides [15]. We perform bit error ratio measurements and achieve error free operation with a penalty of ~ 4.3dB at $10^{-9}$ BER. Our results confirm that free-carrier dynamics, whilst they do adversely influence device performance, do not pose a barrier to device operation at extremely high bit rates. The high bit rate operation of this logic gate chip is a result of the ultrafast response of the Kerr nonlinearity together with the broad phase-matching achievable in this device.

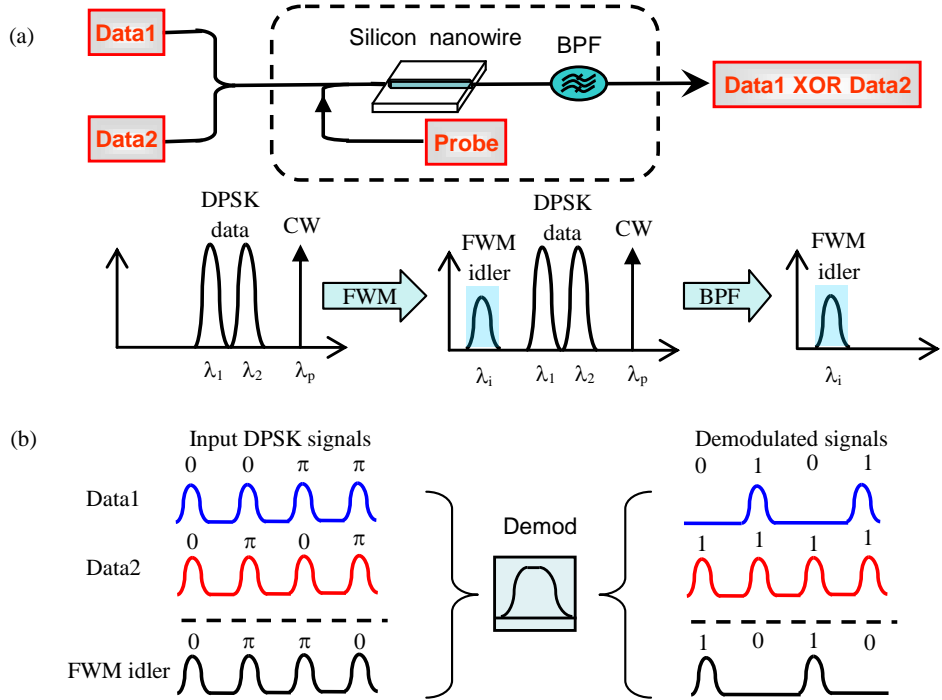

Fig.1 (a) Principle of operation of an all-optical XOR logic gate based on four-wave mixing (FWM) and (b) an illustration of the relationship of the phases between the two DPSK input signals and the output of the XOR gate, the idler produced by FWM between the input signals.

## 2. Theory

Figure 1 (a) shows the principle of an all-optical XOR logic gate based on FWM [9,15]. Two high bit rate DPSK data streams, centered at wavelengths $\lambda_1$ and $\lambda_2$ with phases (reflecting the data "1"s and "0"s) of $\varphi_1$ and $\varphi_2$, respectively, are co-propagated with a third CW signal at $\lambda_p$ (with a constant phase $\varphi_p$) through a silicon nanowire. Four wave mixing between coinciding data streams and the CW probe generates idler pulses at a wavelength of $\lambda_i$ with a phase given by $\varphi_i = \varphi_1+\varphi_2 - \varphi_p$ [10]. The idler is then filtered by an optical band pass filter and directed into a one-bit-delay interferometer, followed by a receiver. Note that in principle the XOR output signal need not necessarily be detected but can simply be re-routed in optical form without electronic conversion or regeneration. Figure 1(b) shows the phase of the filtered FWM idler at the output of the silicon chip for all possible states of the modulation phase of the two DPSK signal inputs. Due to the phase periodicity, a phase of "0" is equivalent to "2π". Thus, the phase of the FWM idler is "π" if the two input data have different modulation phases and "0" if they are in phase. The truth table of a two-input XOR gate in terms of the phase of the FWM idler is shown in the left-hand side of Figure 1 (b) while the demodulated signals are shown in the right-hand side of Figure 1(b) to verify the integrity of the XOR logic function. Note that Figure 1(b) shows the demodulated signals from the destructive port of the DPSK demodulator.

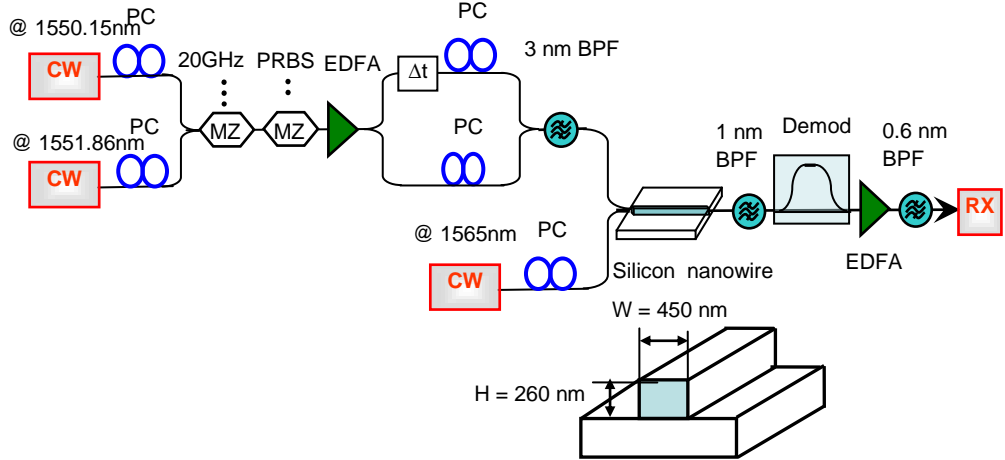

Fig.2 Experimental setup for the XOR logic gate for 40Gb/s (33% RZ) DPSK signals based on a silicon nanowire. MZ = Mach-Zehnder modulator, PC = polarization controller.

### 3. Experiment

The device is a 5mm long silicon-on-inslulator (SOI) "nanowire" with dimensions 450nm wide x 260nm thick (Figure 2), fabricated by electron-beam lithography and reactive ion etching. The waveguides were coated with SU8, and contained "inverse" taper regions to reduce coupling losses. The measured propagation loss was ∼ 3dB/cm for both TE and TM polarizations and coupling losses were 5dB/facet for TM and TE polarizations, achieved via lensed fiber tapers with nanopositioning stages, resulting in a total insertion loss of ∼ 11.5dB (for either polarization). The total dispersion (waveguide plus material) is anomalous and roughly constant at ∼ +500ps/nm/km over the C-band for TE and for TM it is normal and varies between ∼ -8,000 ps/nm/km and ∼ -16,000 ps/nm/km. We therefore used TE polarization for these experiments to ensure the phase-matching required for FWM.

The experimental setup for demonstrating an all-optical XOR logic gate for a 40Gb/s DPSK signal is shown in Figure 2. Two 40 Gb/s RZ-DPSK signals were generated from two CW laser sources centered at $\lambda = 1550.15$ nm and $\lambda = 1551.86$ nm using a pair of Mach-Zehnder electro-optic modulators. The first MZ modulator carved 40 GHz RZ pulses with a 33% duty-cycle, while the second one encoded DPSK data on the pulses at 40 Gb/s with a $2^7-1$ pseudorandom bit sequence (PRBS). The two DPSK data streams were amplified by an erbium-doped fiber amplifier (EDFA) and then separated by an arrayed waveguide grating demultiplexer. An optical delay line (ΔT) created delays between the two data streams. The average power of each signal was 30 mW (incident, or ∼9 mW in the waveguide) for a combined average incident launch power of 60 mW. The CW probe was centered at 1565 nm with an incident average power of 30 mW. The (TE polarized) signal and pump were combined with a 80:20 coupler before launching into the waveguide with polarizations aligned via polarization controllers (PC). The XOR product was then extracted using a 1-nm optical BPF and demodulated by a delay line interferometer before detection with a 40 Gb/s receiver (RX). Note that the destructive interference port of a DPSK demodulator was used in our experiments. An EDFA and a 0.6-nm BPF were used at the receiver end to improve the optical signal-to-noise ratio (OSNR) of the detected signals. We performed system measurements in order to evaluate the transmission bit error ratio (BER) and system penalty of the device.

### 4. Results and discussion

The optical spectra at the waveguide output (Figure 3) shows the probe and signal spectra (separated by ∼ 15 nm) as well as the XOR idler spectrum at 1537nm generated by FWM, showing a conversion efficiency of ∼ -25dB. Also visible to either side of the XOR idler peak are the spectra (at 1539nm and 1536nm) of the two idlers resulting from FWM between the individual DPSK signals and the CW probe. These do not contain cross-terms between the two DPSK signal inputs and so are of no interest in terms of logic operation and are removed by the filter. The side lobes visible around both the two input DPSK signals and the CW probe are a result of FWM between the DPSK signals and cross phase modulation of the CW probe from the RZ signals, and again they are not relevant to the logic operation and do not adversely affect performance.

Figure 4(a) shows the demodulated waveforms of the two input DPSK data streams along with the XOR signal, all extracted by tunable band-pass optical filters followed by an optical sampling oscilloscope (Picosolve). It is clearly observed that the phase of the FWM idler accurately reflects the XOR logic function of two input signals. The corresponding optimized eye diagram is shown in Figure 4(b). All traces were measured with a 65GHz detector and sampling oscilloscope. The eye diagram of the 40 Gb/s demodulated XOR signal is clearly open, highlighting the effective all-optical XOR operation. Finally, the performance of this chip-based XOR logic function was confirmed with bit error-rate (BER) measurements. Figure 5 shows the system penalty measurements indicating a penalty, relative to back-to-back operation, of ~ 3.0dB and ~ 4.3dB at $10^{-9}$ BER.

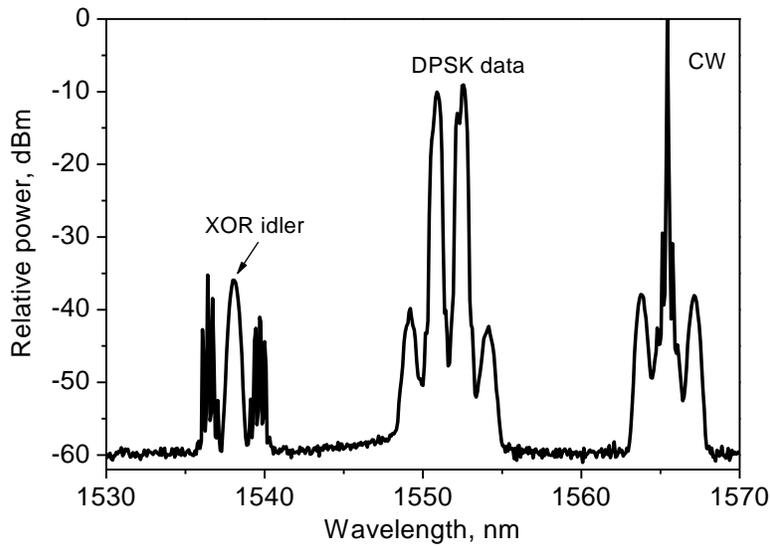

Fig.3 Output optical spectrum measured on an OSA.

The error-free operation of our device clearly demonstrates that free-carriers are not a barrier to effective device operation at these data rates. This is not unexpected since the error-free wavelength conversion via FWM in silicon nanowires has been demonstrated at data rates from 40Gb/s to 1.28Tb/s [23-26]. However, it is clear that reducing free carrier effects will increase the device efficiency and reduce the operating power requirements. We also anticipate a significant reduction in both the system penalty and operating pump power by reducing coupling and propagation losses as well as optimizing the waveguide dispersion closer to the ideal for FWM, namely, small but anomalous. Note that the CW probe, DPSK signals and idler span the entire C-band in these experiments and so we expect efficient device performance across the C-band. All of these factors point to silicon all-optical logic devices based on the ultrafast response of the Kerr nonlinearity being able to operate at substantially higher bit rates.

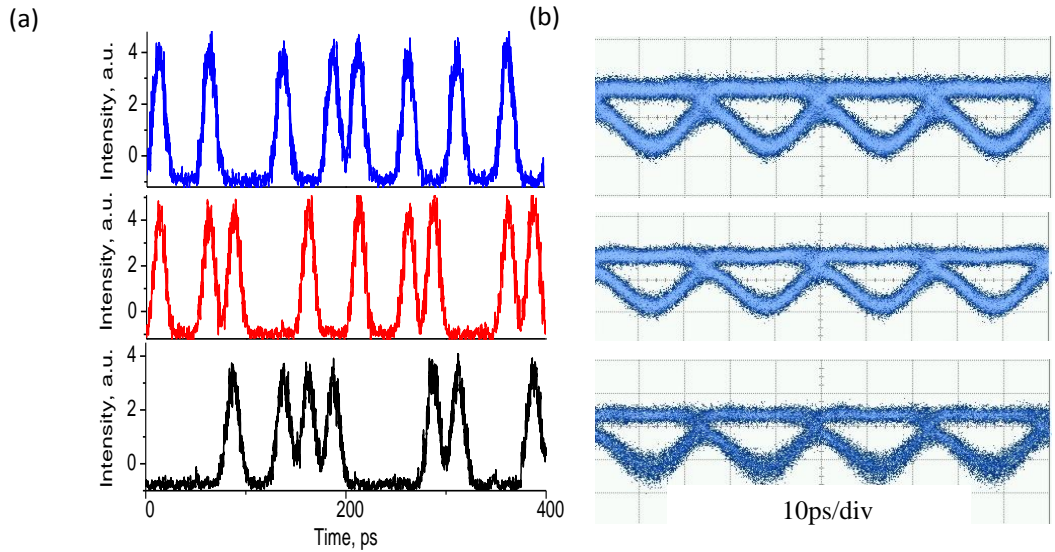

Fig.4 Pulse traces (a) and eye diagrams (b) corresponding to (top) the 40Gb/s input DPSK signal at λ = 1550.18 nm, (middle) the 40Gb/s DPSK input signal at λ = 1551.86 nm and (bottom) the output 40Gb/s idler at λ = 1537nm, representing the XOR logic output.

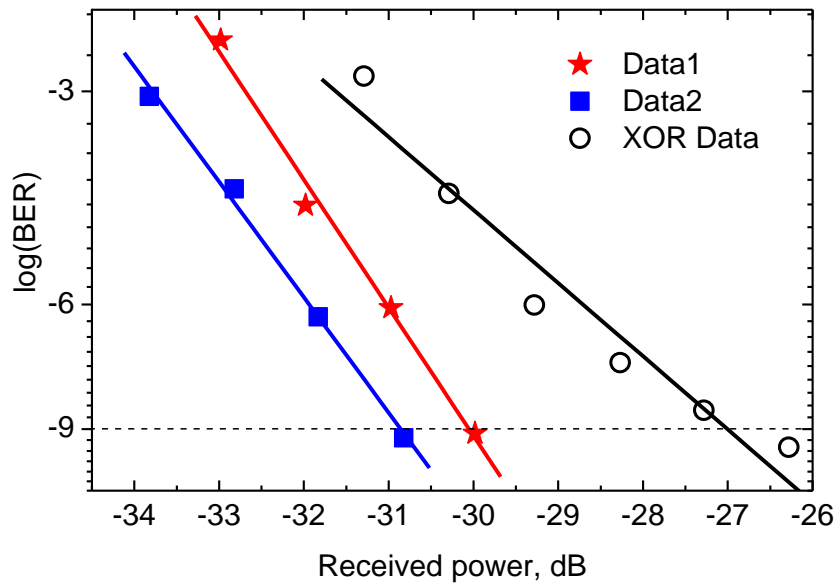

Fig.5 BER measurements for the signal (blue and red lines) and the idler (black line) generated by a 30mW signal and a 30mW probe. The relative power penalty between the curves at a BER of $10^{-9}$ is 3.0dB and 4.3dB, respectively.

## 5. Conclusions

We demonstrate an all-optical XOR logic function operating on a DPSK PRBS data stream at 40Gb/s (33% Return-to-Zero) in the C-band via FWM in a silicon nanowire. We perform bit error ratio measurements and achieve error free operation with a penalty of 3.0dB and 4.3dB at $10^{-9}$ BER. This device is expected to operate at even higher data rates owing to its broad phase-matching and the ultrafast response of the Kerr nonlinearity.

## Acknowledgements

This research was supported by the Australian Research Council (ARC) Centers of Excellence (COE), Discovery Projects and Federation Fellowship programs.